\documentclass{aa}  

\usepackage{graphicx}
\usepackage{txfonts}
\usepackage{amsmath}
\newcommand{\formaldehyde}{$\mathrm{H_2CO}$}

\newcommand{\htwo}{$\mathrm{H_2}$}
\newcommand{\gstate}{$ 1_{10}-1_{11}$}
\newcommand{\fstate}{$ 2_{11}-2_{12}$}
\newcommand{\sstate}{$ 3_{12}-3_{13}$}
\newcommand{\gsstate}{$ 3_{13}-1_{11}$}
\newcommand{\zerostate}{$ 1_{11}$}

\newcommand{\taug}{$\tau_{4.8}$}

\newcommand{\nhtwo}{$n_{\mathrm{H_2}}$}
\newcommand{\tk}{$T_k$}
\newcommand{\td}{$T_d$}

\begin{document}

   \title{Revisiting the formaldehyde masers}


   \author{D.J. van der Walt and L.L. Mfulwane
          }

   \institute{Centre for Space Research, North-West University, Potchefstroom 2520, South Africa\\
              \email{johan.vanderwalt@nwu.ac.za}
             }

   \date{Received September 15, 1996; accepted March 16, 1997}

 
  \abstract
  {The 4.8 GHz
  formaldehyde (\formaldehyde{}) masers are one of a number of rare types of molecular
  masers in the Galaxy. There still is not agreement on the mechanism responsible for the
  inversion of the \gstate{} transition and the conditions under which an inversion can
  occur, and therefore how to interpret the masers.  }
      {
        The aim of the present calculations is to explore a
        larger region of parameter space to improve on our previous calculations,
        thereby to better understand the range of physical
        conditions under which an inversion of the \gstate{} transition occurs. We also aim
        to understand recently published results that  
        \formaldehyde{} masers are radiatively pumped.
      }
      {
        We solve the rate equations of the first 40
        rotational levels of o-\formaldehyde{} using a fourth-order Runge-Kutta method. We
        consider gas kinetic temperatures between 10 K and 300 K, \htwo{} densities between
        $10^4~\mathrm{cm^{-3}}$ and $10^6~\mathrm{cm^{-3}}$, and a number of different dust
        temperatures and grey-body spectral energy density distributions.
      }
      { We show that when using a black body radiation field the inversion of any
        transition will disappear as the kinetic temperature approaches the black-body
        radiation temperature since the system, consisting of the gas and radiation field,
        approaches thermodynamic equilibrium. Using a grey-body dust radiation field
        appropriate for Arp 220 we find that none of \gstate{}, \fstate{}, and \sstate{}
        transitions are inverted for kinetic temperatures less than 100 K. Our
        calculations also show that in theory the \gstate{} transition can be inverted
        over a large region of explored parameter space in the presence of an external
        far-infrared radiation field. Limiting the abundance of \formaldehyde{} to less
        than $10^{-5}$, however, reduces the region where an inversion occurs to \htwo{}
        densities $\gtrsim 10^5~\mathrm{cm^{-3}}$ and kinetic temperatures $\gtrsim$ 100
        K. We propose a pumping scheme for the \formaldehyde{} masers which can explain
        why collisions play a central role in inverting the \gstate{} transition, and
        therefore why an external radiation field alone does not lead to an inversion.  }
      {
        Collisions are an essential mechanism for the inversion of the \gstate{}
        transition. Our results suggest that 4.8 GHz \formaldehyde{} megamasers are
        associated with hot and dense gas typical of high mass star forming regions rather
        than with cold material. Although limiting the
        \formaldehyde{} abundance to less than $10^{-5}$ significantly reduces the region
        in parameter space where the \gstate{} is inverted, it still is not clear whether
        this is the only reason why these masers are so rare.
      }


   \maketitle
%

\section{Introduction}
\label{introduction}

The 4.8 GHz ortho-formaldehyde (hereafter  \formaldehyde{}) \gstate{}
masers belong to a small group of rare astrophysical masers. At present only eight
Galactic high mass star forming regions with associated 4.8 GHz \formaldehyde{} masers are
known \citep{Araya2008, Ginsburg2015},  compared with the more than 900 6.7
GHz methanol masers associated with high mass star forming regions detected in the
Methanol Multibeam survey \citep{Green2017}. Since the \formaldehyde{} masers seem to be
exclusively associated with high mass star forming regions \citep{Araya2015}, the question
arises of what is special about these star forming regions with associated
\formaldehyde{} masers compared to the majority of high mass star forming regions without
associated \formaldehyde{} masers.

High luminosity 4.8 GHz megamaser emission has also been detected from the nuclear regions
of three ultraluminous infrared (starburst) galaxies \citep{Baan1986, Araya2004,
  Araya2008, Baan2017}. Analogously to the question about the excitation of the 4.8 GHz
masers in a few Galactic high mass star forming regions, the same question arises for the
extragalactic \formaldehyde{} megamasers. All three known \formaldehyde{} megamaser
sources are also OH megamaser sources, which obviously raises the question of whether the two
types of masers are excited in the same way.

Since the paper by \citet{Boland1981}, the question of the pumping of  \formaldehyde{}
masers has regularly been raised over a period of about four decades. Recently, a number of
authors  \citep[see e.g.][]{Ginsburg2015,Lu2019} have remarked that the issue of the pumping
of the 4.8 GHz \formaldehyde{} masers is still  unclear, which makes it difficult to
interpret their observations. Further uncertainty also followed from the statement by
\citet{Baan2017} that the masers are radiatively pumped, while \citet{vanderwalt2014}
showed that an inversion of the \gstate{} transition can be achieved without the presence
of an external far-infrared radiation field. \citet{vanderwalt2014} also found that no
inversion could be found if collisions are switched off.

In view of the  uncertainties in the pumping of the \formaldehyde{} masers, we expanded
the calculations of \citet{vanderwalt2014} to better cover the parameter space and thereby to
obtain a  more complete picture of the inversion of the \gstate{} transition. We
also present the results of calculations aimed at understanding the results obtained with
RADEX by \citet{Baan2017} and whether inversion of the \gstate{} transition can indeed be
interpreted as being due to radiative pumping (by a black-body radiation field) as claimed
by these authors. Following from the results, we propose a pumping scheme for the
\formaldehyde{} masers that explains why collisions play a central role in the inversion
of the \gstate{} transition. For the present purposes we do not consider the effect of
beaming, but focus only on the physical conditions under which an inversion can occur.

\section{Molecular data}

Formaldehyde is  a near prolate symmetric top molecule with the dipole moment along the
C--O axis (the A-axis). The energy levels are characterized by three quantum numbers, $J$,
$K_a$, and $K_c$, with $J$ the total angular momentum, $K_a$ the projection of $J$ on the A-axis (the symmetry axis for a limiting prolate symmetric top), and $K_c$ the projection of
$J$ on the C-axis (the symmetry axis for a limiting oblate symmetric top). In the case of
o-\formaldehyde{}, where  the nuclear spins of the two hydrogen atoms are parallel,
$K_a$ is odd. The slight asymmetry of the molecule causes each rotational level to be
split into two energy levels (known as K-doublets) characterized by different values of
$K_c$ such that $K_c$ is even for the higher level and odd for the lower level.

\begin{figure}
  \centering
  \includegraphics[scale = 0.5]{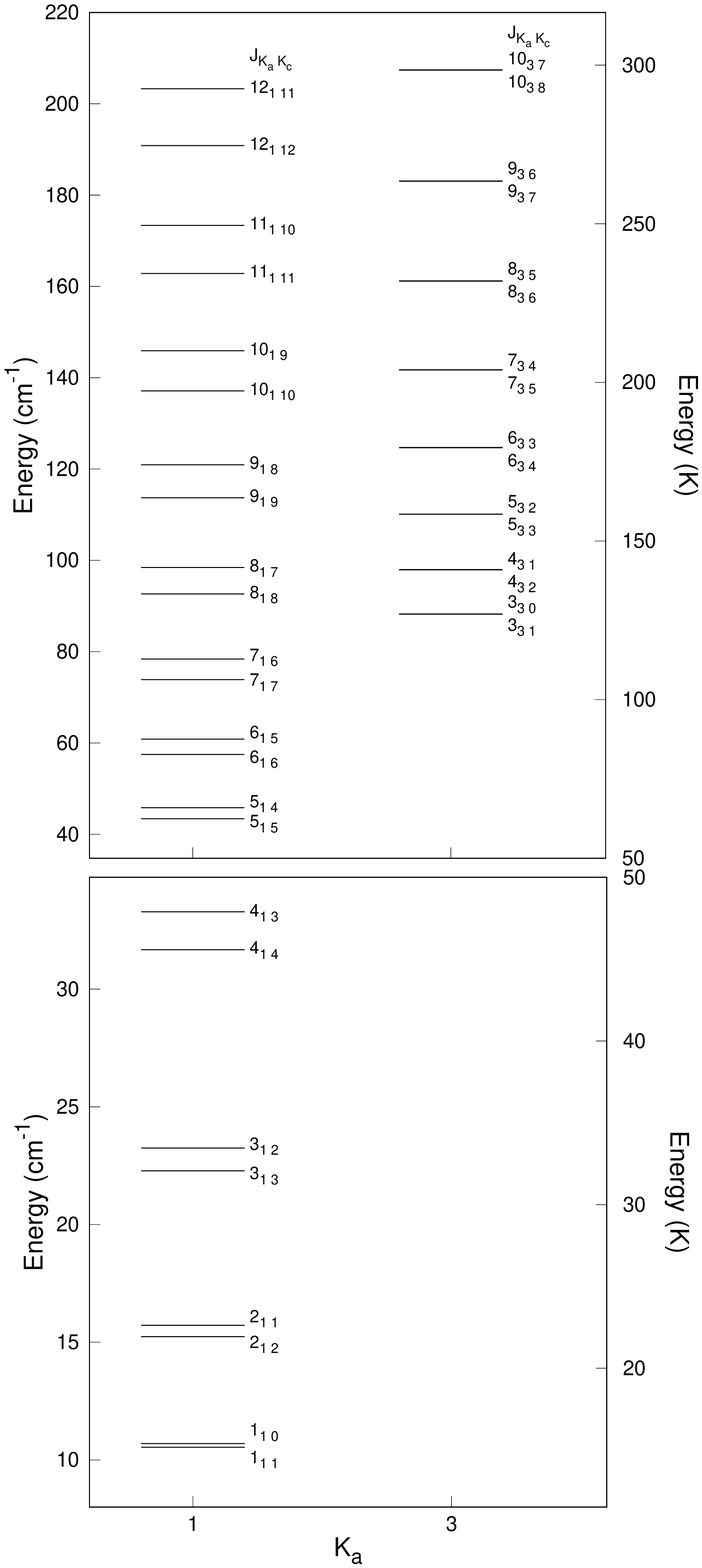}
  \caption{Energy level diagram for the first 40 levels of ortho-formaldehyde. {\it Lower
      panel:} The  first eight levels with E < 50 K.  {\it Upper panel:} The levels
    with E > 50 K. The energy scales are different on the y-axes for the lower
    and upper panels to better separate the doublet states at lower energies.}
  \label{fig:elevels}
\end{figure}

The  energy level diagram for the first 40 levels of o-\formaldehyde{} used in this work
is shown in Fig.\,\ref{fig:elevels} for $K_a = 1,3$. The energy levels and Einstein A
coefficients for allowed radiative transitions were taken from the Leiden Atomic and
Molecular Database \citep{Schoier2005}. The coefficients for collisions of
o-\formaldehyde{} with o-$\mathrm{H_2}$ and p-$\mathrm{H_2}$, as calculated by
\citet{Wiesenfeld2013}, were also obtained from the Leiden Atomic and Molecular
Database. In all calculations the ratio of o-$\mathrm{H_2}$ to p-$\mathrm{H_2}$ was that
for thermodynamic equilibrium at the specific kinetic temperature. For kinetic
temperatures between 100 K and 300 K, a linear interpolation was used to produce collision
rate coefficients for 10 K intervals instead of the 20 K intervals  listed in the Leiden
Atomic and Molecular Database.

Formaldehyde has six vibrational normal modes with the lowest at an energy of 1167
$\mathrm{cm^{-1}}$ $\equiv$ 1679 K \citep{Nikitin2021}. This is significantly higher than
the excitation energies associated with high mass star forming regions and it is therefore
not necessary to consider excitation to the vibrational states.

\section{Numerical method}

The level populations were found by solving the well-known set of rate equations, which in
the escape probability approach is given by
\begin{eqnarray}\nonumber
\frac{dN_i}{dt} & = & \sum_{j < i}[(-N_i + (\frac{g_i}{g_j}N_j -
  N_i)W\mathcal{N}_{ij})\beta_{ij}A_{ij} \\  
\nonumber & &  + C_{ij}(N_j\frac{g_i}{g_j}e^{-E_{ij}/{kT}} - N_i)]\\
\nonumber  & & +  \sum_{j > i}[(N_j + (N_j  -
\frac{g_j}{g_i}N_i)W\mathcal{N}_{ji})\beta_{ji}A_{ji} \\ 
& & + C_{ji}(N_j - N_i\frac{g_j}{g_i}e^{-E_{ji}/{kT}})]
\label{eq:rate}
.\end{eqnarray}
Here $N_i$ is the number density in level $i$, $g_i$ the statistical weight of level $i$,
$A_{ij}$ the Einstein A coefficient for spontaneous emission between levels $i$ and $j$;
$W$ is the geometric dilution factor for an external radiation field 
with a spectral energy distribution (SED) given by
\begin{equation}
  F_{\nu}(T_d) = [1 - e^{-(\nu/\nu_0)^p}]B_\nu(T_d)
  \label{eq:dustsed}
,\end{equation}
and $\mathcal{N}_{ij}$ the photon occupation number for this field at frequency
$\nu_{ij}$;  $C_{ij} = n_{H_2} K_{ij}$ is the collision rate with $n_{H_2}$ the \htwo{}
number density and $K_{ij}$ the collision rate coefficient. The fiducial frequency
  $\nu_0$ was taken as $3\times10^{12}$ Hz; $\beta_{ij}$ is the escape probability for
which we used the expression for the large velocity gradient approximation 

\begin{equation}
  \beta_{ij} = \frac{1 - e^{-\tau_{ij}}}{\tau_{ij}}
  \label{eq:escprob}
  ,\end{equation}
with the optical depth given by 
\begin{equation}
  \tau_{ij} =
\frac{A_{ij}}{8\pi}\left(\frac{c}{\nu}\right)^3\left(\frac{g_i}{g_j}x_j -
x_i\right)\frac{N_{col}}{\Delta \varv}.
\label{eq:tau}
\end{equation}
The quantities $x_i$ and $x_j$ are the fractional number densities of molecules in the upper level $i$
and lower level $j$, respectively; $N_{col}$ is the \formaldehyde{} column density and
$\Delta \varv$ the line width. The ratio $N_{col}/\Delta \varv$ is the specific column
density.

The rate equations were supplemented with the particle number conservation requirement
\begin{equation}
N_{tot} = Xn_{H_2} = \sum_i N_i
,\end{equation}
with $X$ the abundance of \formaldehyde{} relative to \htwo{}.

We solve the rate equations using a fourth-order Runge-Kutta method. The initial
distribution of the level populations was a Boltzmann distribution with temperature equal
to the average of the dust and kinetic temperatures.  For a given \nhtwo{} the calculation
started with a small \formaldehyde{} specific column density ($10^5~\mathrm{cm^{-3}\,s}$)
such that the system is completely optically thin in all transitions. An equilibrium
solution for this initial specific column density was then found by letting the system
evolve in time steps of 5 seconds until the convergence condition
$|N_i(t_{j+1})-N_i(t_j)|/N_i(t_j) < 10^{-6}$ is reached for all levels.  The specific
column density is then increased by a small amount (0.001 dex); the equilibrium solution
for the previous value of the specific column density is used as the initial
distribution. The process is repeated until a specific column density of $5 \times
10^{13}~\mathrm{cm^{-3}\,s}$ is reached or is stopped if \taug{} > 5 before this specific
column density is reached. If there is an inversion in the range of specific column
densities, the maximum negative optical depth and the corresponding specific column
density are recorded. If there is no inversion \taug{} = 0.

\section{Results}

\subsection{The \citet{Baan2017} case}
\label{sec:baan}
We consider first Fig.\,19 in \citet{Baan2017}  which  shows that inversion
of the \gstate{} and \fstate{} transitions of \formaldehyde{} can be achieved when locally
the gas kinetic temperature is lower than the dust radiative temperature. Understanding the
behaviour of the optical depth for different kinetic temperatures  is essential before making any conclusions based on the shown behaviour.

\begin{figure}
  \centering
  \includegraphics[scale=0.85]{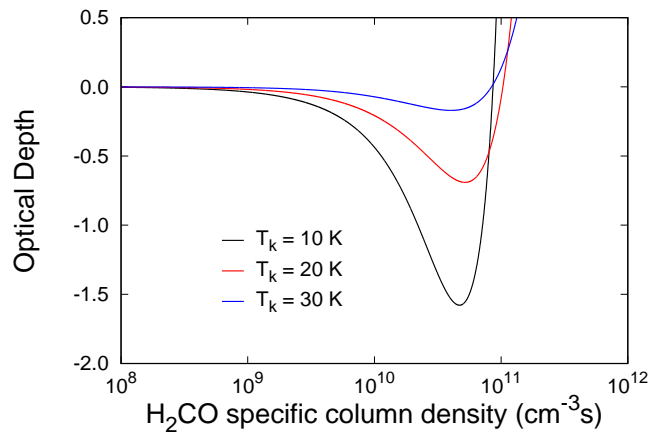}
  \caption{4.8 GHz optical depth as a function of the \formaldehyde{} specific column
    density for \nhtwo = $10^4~\mathrm{cm^{-3}}$ and \td{} = 50 K. }
  \label{fig:baancase}
\end{figure}

\citet{Baan2017} obtained the results shown in their Fig.\,19 using the online version of
Radex \citep{vandertak2007}. In particular, it should be noted that the online version
of Radex uses an unshielded black-body radiation field \citep{vanlangevelde2008} for the
dust radiation. In Fig.\,\ref{fig:baancase} we show the dependence of the optical depth on
the \formaldehyde{} specific column density for the \gstate{} transition for \tk{} = 10,
20, 30 K; \td{} = 50 K (undiluted black body); and \nhtwo{} = $10^4~\mathrm{cm^{-3}}$
obtained with the above-described numerical method to solve the rate equations. Although
there are quantitative differences in the optical depths when compared with Fig.\,19 of
\citet{Baan2017}, the behaviour for the different values of the kinetic temperature is the
same, which means that  the inversion is largest for \tk{} = 10 K and becomes smaller as the kinetic
temperature increases to approach the dust temperature.

\begin{figure}
  \centering
  \includegraphics[scale = 0.75]{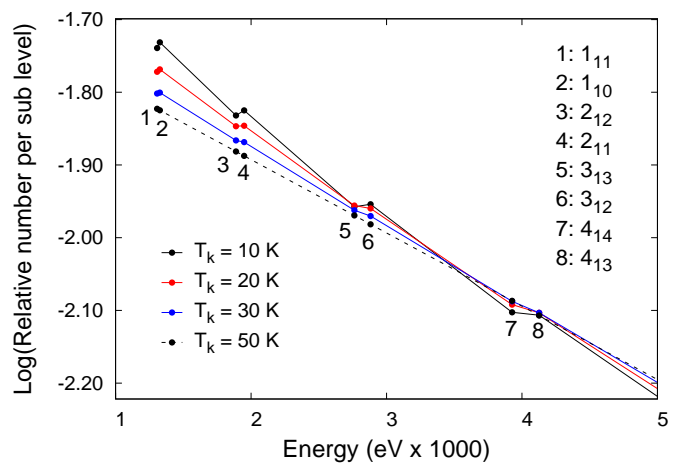}
  \caption{Comparison of the level populations for levels $J_{K_a,K_c}$ from
    $\mathrm{1_{1,1}}$ to $\mathrm{4_{1,3}}$ for different kinetic temperatures when
    \nhtwo = $10^4~\mathrm{cm^{-3}}$ and the external radiation field is a black body with
    \td{} = 50 K. The levels are numbered from 1 to 8 just below the case for \td{} = 50 K
  }
  \label{fig:baanlevpops}
\end{figure}

Figure \ref{fig:baancase} therefore confirms the result of \citet{Baan2017} that under the
above-stated conditions the \gstate{} transition shows an inversion.  We note that since
the radiation field used is that of an undiluted black body, increasing the kinetic
temperature to approach the radiation temperature brings the system closer to
thermodynamic equilibrium. Thus, when \tk{} = \td{} the level populations follow a
Boltzmann distribution, in which case a population inversion is not possible.  We
illustrate this further in Fig.\,\ref{fig:baanlevpops} where we compare the level
populations for the lower energy levels where inversions occur, when \tk{} = 10, 20, 30,
50 K. The inversions for the different transitions are largest for \tk{} = 10 K and are
seen to become smaller as \tk{} approaches 50 K. For \tk{} = \td{} = 50 K there is no
inversion and, consistent with thermodynamic equilibrium, the level populations follow a
Boltzmann distribution with a temperature of 50 K.

\begin{figure}
  \centering
  \includegraphics[scale=0.6]{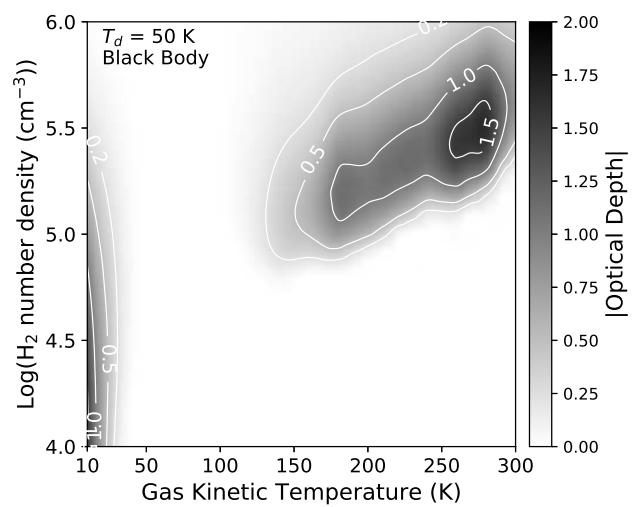}
  \caption{Variation in the 4.8 GHz optical depth in the \nhtwo{}-\tk{} plane when the
    spectral energy distribution (SED)
    of the external emission is given by an undiluted black body at 50 K. \taug{} = 0
    means that there is no inversion.}
  \label{fig:taubbdust50K}
\end{figure}

Still within the framework of the scenario proposed by \citet{Baan2017}, we further
examined the behaviour of \taug{} in the \nhtwo{}-\tk{} plane for $\mathrm{10~K} \le T_k
\le \mathrm{300~K}$, $10^4 ~\mathrm{cm^{-3}}\le n_{H_2} \le 10^6~\mathrm{cm^{-3}}$, and
when dust emission is that of an undiluted black body at 50 K.  The result is shown in
Fig.\,\ref{fig:taubbdust50K}. It is seen that there are two distinct regions in the
\nhtwo{}-\tk{} plane where the \gstate{} transition is inverted. The first is a small
region with \tk{} < 40 K and which corresponds with the results shown in
Fig.\,\ref{fig:baancase}. The second is a much larger region with \tk{} $\gtrsim$ 130 K
and \nhtwo{} $\gtrsim 10^{4.8}~\mathrm{cm^{-3}}$.  The absence of an inversion for kinetic
temperatures around 50 K is simply a manifestation of the system being close to
thermodynamic equilibrium. As we show below, the second region corresponds to the
region where inversion is due to collisions only (i.e. without the presence of an external
dust infrared radiation field).

\begin{figure}
  \centering \includegraphics[scale=0.68]{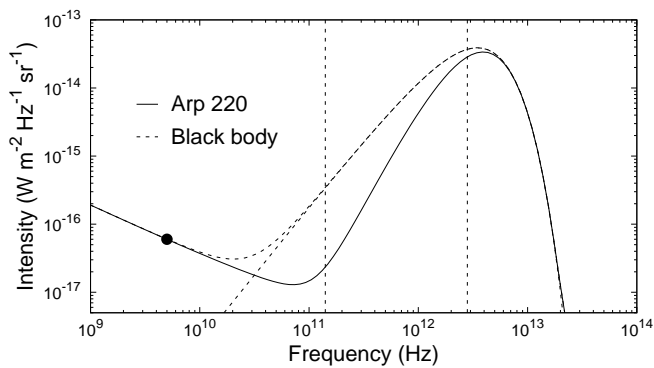}
  \caption{Comparison of a black-body SED with the
    radio--to--far-infrared SED for Arp 220 \citep{Yun2002}. The
    left vertical dashed line is at 140.8 GHz, which is the  frequency for the
    \gsstate{} transition. The right vertical dashed line is at 2808 GHz, which is the
    highest transition frequency for the 40 levels. }
  \label{fig:arp220}
\end{figure}

\begin{figure}
  \centering
  \includegraphics[scale=0.6]{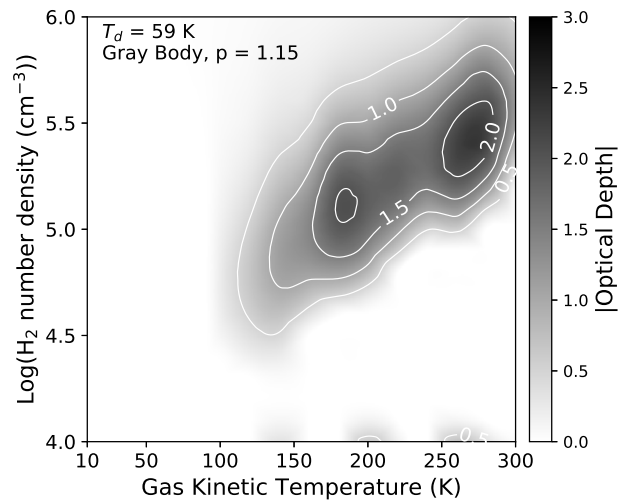}
  \caption
      { Variation of $\lvert\tau_{4.8}\rvert$ in the \nhtwo{}-\tk{} plane when the SED of
        the dust emission is given by $F_{\nu}(T_d) = [1 - e^{-(\nu/\nu_0)^p}]B_\nu(T_d) $
        with $p=1.15$ and $\mathrm{T_d}$ = 59 K as for Arp 220 \citep{Yun2002}. The
        synchrotron and free-free components  shown in Fig.\,\ref{fig:arp220} have also
        been included.}
  \label{fig:tauarpdust}
\end{figure}

In the present case inversion of the \gstate{}, \fstate{}, and \sstate{} transitions has
been achieved using a black-body radiation field for the dust emission. In reality the
dust radiation field is not that of a black body. Staying within the scenario presented by
\citet{Baan2017} that the dust temperature is $\sim$ 50 K, we used the
radio--to--far-infrared SED of Arp 220 derived by \citet{Yun2002}. The SED of the dust
emission is given by Eq.\,\ref{eq:dustsed} with \td{} = 59 K and $p$ = 1.15
\citep{Yun2002} and is compared in Fig.\,\ref{fig:arp220} with that of a black body at the
same temperature. In Fig.\,\ref{fig:tauarpdust} we show the corresponding variation of
$\lvert\tau_{4.8}\rvert$ in the \nhtwo{}-\tk{} plane. Comparison with
Fig.\,\ref{fig:taubbdust50K} shows that while at higher densities and kinetic temperatures
the shape of the region where there is an inversion of the \gstate{} transition is more or
less the same, there is no inversion of the \gstate{} transition for \tk{} $\lesssim$ 100
K in this case.

\subsection{Expanding the calculations of \citet{vanderwalt2014}}

\citet{vanderwalt2014} presented results of the inversion of the \gstate{} transition for
a limited number of selected combinations of \tk{}, \nhtwo{}, and \td{} which represented only
small parts of the parameter space. We now turn to presenting somewhat more general results
covering 10 K $\le$ \tk{} $\le$ 300 K and $10^4 ~\mathrm{cm^{-3}}\le n_{H_2} \le
10^6~\mathrm{cm^{-3}}$, where we consider the effect of having no external radiation field
as a source of excitation and we  investigate  the effect  of an external
radiation field on the inversion of the \gstate{} transition.

\begin{figure}
  \centering
  \includegraphics[scale=0.6]{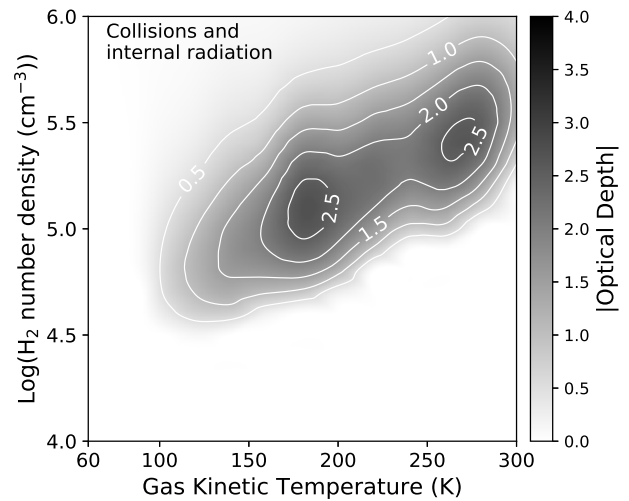}
  \caption{Variation of $\lvert\tau_{4.8}\rvert$ in the $n_{H_2}-T_k$ plane when
    there is no external dust radiation field. Excitation is only through collisions and
    the internal radiation field.}
\label{fig:taucollonly}
\end{figure}

In Fig.\ref{fig:taucollonly} we show the behaviour of $ \lvert \tau_{4.8} \rvert$ in the
$n_{H_2}-T_k$ plane when the external radiation field is switched off (i.e. excitation is
due only to the internal radiation field and collisions). Inversion of the \gstate{}
transition occurs for \tk{} $\gtrsim$ 100 K and \nhtwo{} $\gtrsim
~\mathrm{10^{4.5}~cm^{-3}}$. This result supports the results of \citet{vanderwalt2014}
that the \gstate{} transition can be inverted without the presence of an external
far-infrared radiation field. It does not, however, explain what role collisions play in
the pumping of the masers.

\begin{figure}
  \centering
  \includegraphics[scale=0.57]{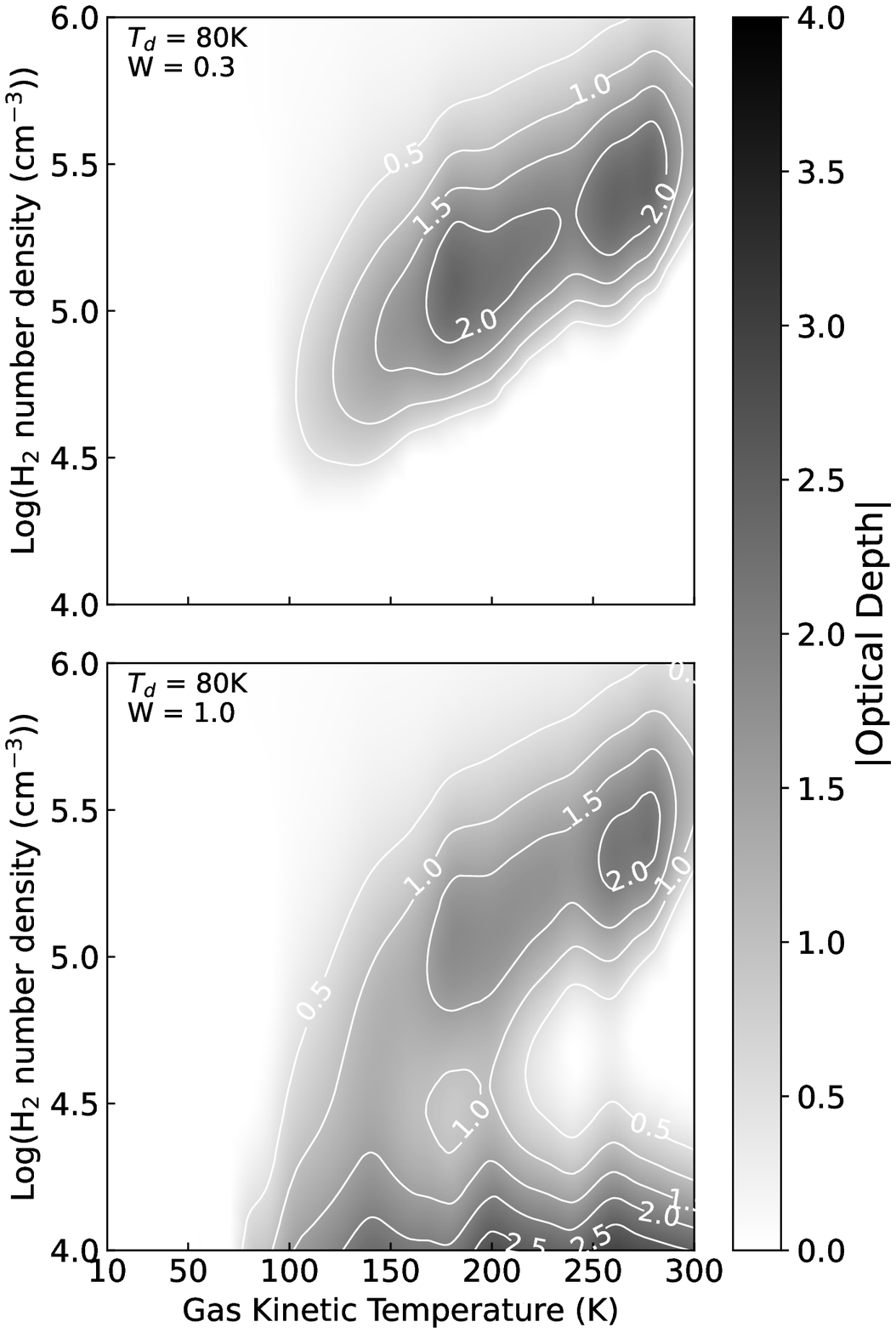}
  \caption
      {
        Variation of $\lvert\tau_{4.8}\rvert$  in the \nhtwo{}-\tk{} plane when the SED
        of the dust emission is given by Eq.\,\ref{eq:dustsed} with $p = 1.15$ and
        $\mathrm{T_d}$ = 80 K. The synchrotron and free-free components   shown in
        Fig.\,\ref{fig:arp220} have also been included. The upper panel is for W = 0.3 and the
        lower panel for W = 1.}
  \label{fig:tautdgb80K}
\end{figure}

\begin{figure}
  \centering
  \includegraphics[scale=0.57]{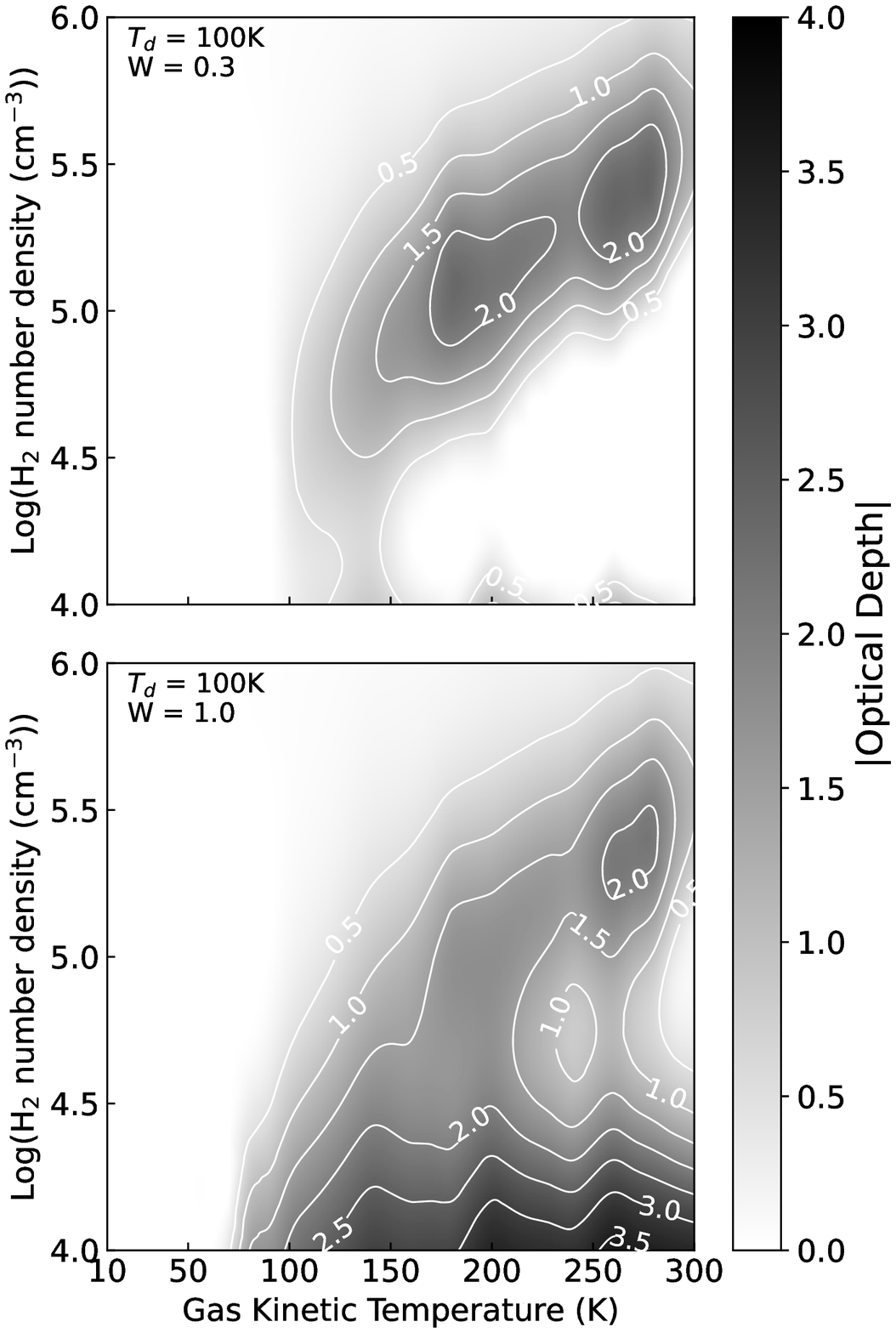}
  \caption
      {
        Variation of $\lvert\tau_{4.8}\rvert$ in the \nhtwo{}-\tk{} plane when the SED
        of the dust emission is given by Eq.\,\ref{eq:dustsed} with $p = 1.15$ and
        $\mathrm{T_d}$ = 100 K. The synchrotron and free-free components   shown in
        Fig.\,\ref{fig:arp220} have also been included. The upper panel is for W = 0.3 and the
        lower panel for W = 1.
      }
  \label{fig:tautdgb100K}
\end{figure}

\begin{figure}
  \centering
  \includegraphics[scale=0.57]{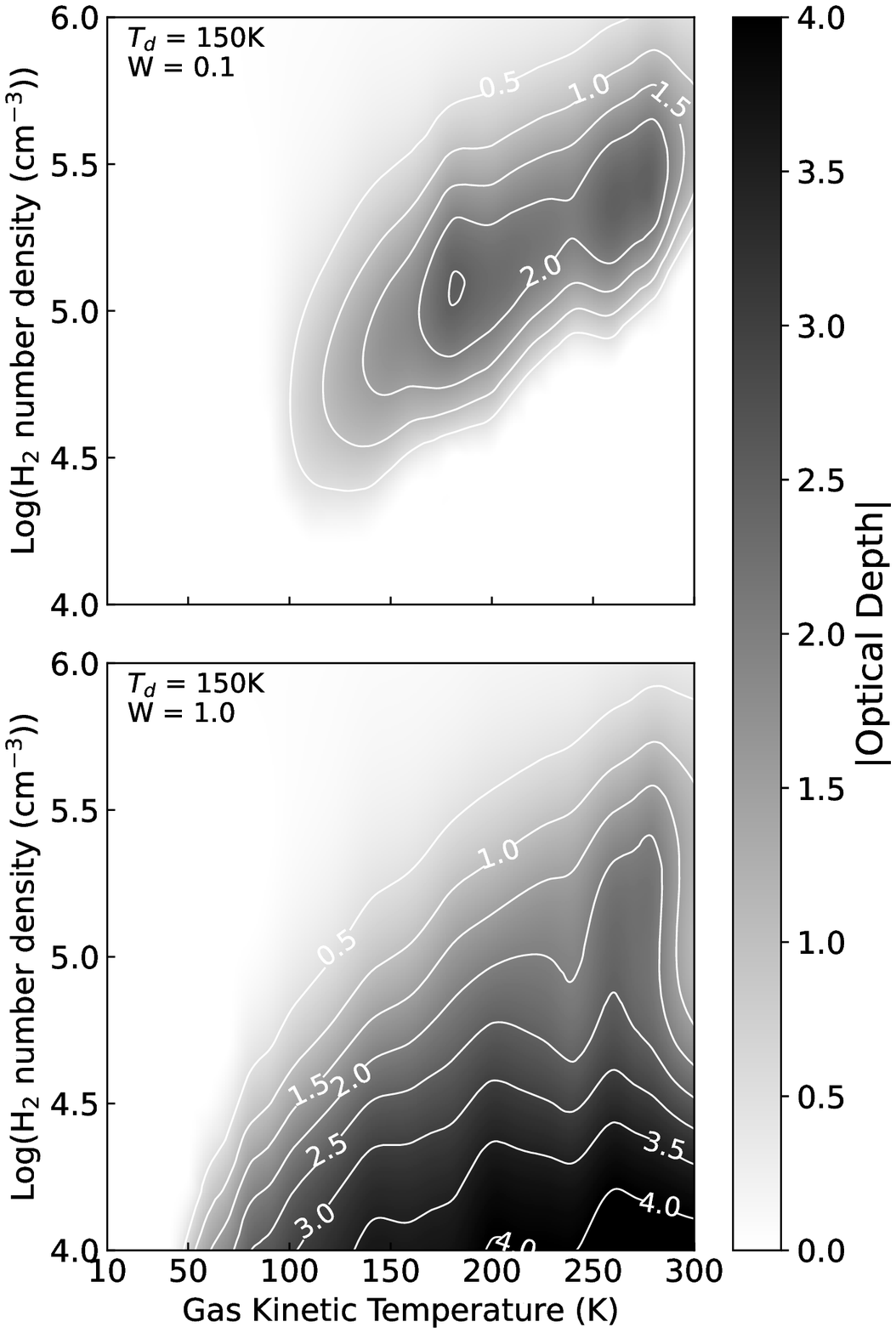}
  \caption
      {
        Variation of $\lvert\tau_{4.8}\rvert$ in the \nhtwo{}-\tk{} plane when the SED
        of the dust emission is given by Eq.\,\ref{eq:dustsed} with $p = 1.15$ and
        $\mathrm{T_d}$ = 150 K. The synchrotron and free-free components   shown in
        Fig.\,\ref{fig:arp220} have also been included. The upper panel is for W = 0.1 and
        the lower panel for W = 1.
      }
  \label{fig:tautdgb150K}
\end{figure}

With regard to the choice of dust and gas kinetic temperatures, it is noted that
\citet{Araya2015} concluded that Galactic \formaldehyde{} masers are exclusive tracers of
high mass star forming regions. Various studies of Galactic high mass star forming regions
have found dust temperatures to be greater than $\sim 50 - 60$ K \citep[see
  e.g.][]{Kraemer2001, Barbosa2016, Lim2019}. This also applies to higher resolution
observations of Arp 220 by  \citet{Wilson2014}, among others, who derived average dust temperatures of
80 K and 197 K for the eastern and western nuclei, respectively. \citet{Downes2007} limits
the intrinsic dust temperature in the western nucleus of Arp 220 to between 90 and 180 K
based on the size and luminosity of the region. \citet{Scoville2015} and
\citet{Scoville2017} derived a dust temperature $>$ 100 K for the western nucleus of Arp
220. We therefore considered three more cases with \td{} = 80, 100, and 150 K to
illustrate the effects of higher dust temperatures and of geometric dilution. While
presenting these results we acknowledge  that the combinations of \td{}, \tk{}, and
\nhtwo{} presented do not necessarily take the thermal balance between dust and gas into
account.

In Figs.\,\ref{fig:tautdgb80K} - \ref{fig:tautdgb150K} we therefore show the behaviour of
$\lvert\tau_{4.8}\rvert$ in the \nhtwo{}-\tk{} plane for \td{} = 80, 100, and 150 K,
respectively. The lower panels in these figures are for the case of an undiluted radiation
field (W = 1 in Eq.\,\ref{eq:dustsed}).  The upper panels in Figs.\,\ref{fig:tautdgb80K}
and \ref{fig:tautdgb100K} are for geometric dilution factors W = 0.3, while  the
upper panel in Fig.\,\ref{fig:tautdgb150K} is for  W = 0.1. The effect of higher dust temperatures is
 interesting in that the region in the \nhtwo{}-\tk{} plane over which the \gstate{}
transition is inverted changes significantly when W = 1. In
Figs.\,\ref{fig:tauarpdust} and \ref{fig:taucollonly} inversion occurs basically for \tk{}
$\gtrsim$ 100 K and \nhtwo{} $\gtrsim~10^{4.5}~\mathrm{cm^{-3}}$, whereas for warmer dust
the inversion occurs at lower \htwo{} densities and also somewhat lower kinetic
temperatures. An undiluted warmer dust radiation field also results in larger values of
$\lvert\tau_{4.8}\rvert$ when compared to the case shown in Fig.\,\ref{fig:taucollonly}. On
the other hand, significant dilution (0.1 - 0.3) of the external radiation field in all
three cases leads to a result  very similar to that when collisions and the
internal radiation field are the only mechanisms for excitation.

Closer inspection of Figs.\,\ref{fig:tautdgb80K} and \ref{fig:tautdgb100K} shows the
presence of faint regularly spaced vertical regions of weak inversion of the \gstate{}
transition. For example,  the top panel of Fig.\,\ref{fig:tautdgb100K} shows that
for W = 0.3 the larger region of inversion connects to lower densities at \tk{} $\sim$
140 K and that regions of weak inversions also start to appear at \tk{} $\sim$ 200 K and
260 K. This results in the wavy nature of the contours for \nhtwo{} $\lesssim~
10^{4.5}\mathrm{cm^{-3}}$ seen in both Figs.\,\ref{fig:tautdgb80K} and
\ref{fig:tautdgb100K}. Test runs with smaller time steps, smaller steps in specific column
density, and  different convergence criteria were performed to check if this behaviour is
due to the numerical method used. In all these cases the wavy behaviour of the optical
depth was present. The exact reason for this behaviour of the optical depth is not clear
and has not been investigated further.

\begin{figure}
  \centering
  \includegraphics[scale=0.57]{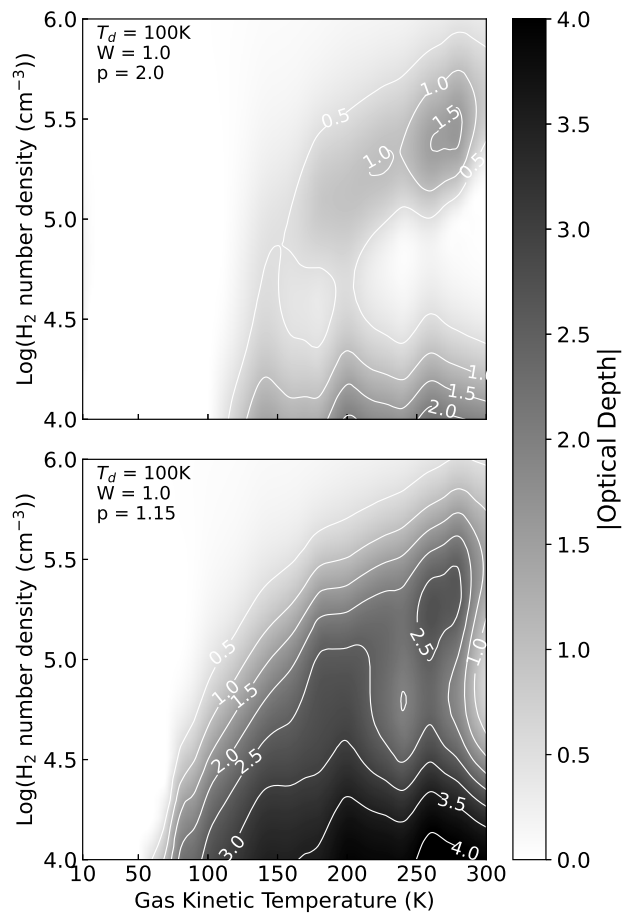}
  \caption
      {
        Comparison of the variation of $\lvert\tau_{4.8}\rvert$ in the \nhtwo{}-\tk{}
        plane when the SED of the dust emission is given by $F_{\nu}(T_d) = [1 -
          e^{-(\nu/\nu_0)^p}]B_\nu(T_d) $ with $p=0.5$ in the upper panel, $p=2.0$ in the
        lower panel, and \td{} = 100 K in both cases.
      }
  \label{fig:tautdgb100Kbeta}
\end{figure}

In addition to  the dilution factor, we finally consider the effect of the shape of the SED of
the dust emission. In Fig.\,\ref{fig:tautdgb100Kbeta} we show in the upper panel the
variation of $\lvert\tau_{4.8}\rvert$ for \td{} = 100 K when $p = 0.5$ in
Eq.\,\ref{eq:dustsed} and in the lower panel when $p = 2.0$. These two cases should be
compared with the lower panel of Fig.\,\ref{fig:tautdgb100K} for which $p = 1.15$. It can be
seen that although the region over which the \gstate{} transition is inverted is more or
less the same for the three values of $p$, $\lvert\tau_{4.8}\rvert$ is smaller over the
whole region for $p = 0.5$ and larger when $p = 2.0$. In the Rayleigh-Jeans limit and for
$(\nu/\nu_0)^p \ll 1$ in Eq.\,\ref{eq:dustsed}, $F_\nu(T_d) \propto \nu^{2+p}$. When $p =
0$, $F_\nu(T_d) = 0.63B_\nu(T_d) \propto \nu^2$, which is the flattest possible SED in the
Rayleigh-Jeans regime. Except for the factor of 0.63, the result presented in
Fig.\,\ref{fig:taubbdust50K} corresponds to the case when $p = 0$.  Therefore, considering
Figs.\,\ref{fig:taubbdust50K}, \ref{fig:tautdgb100K}, and \ref{fig:tautdgb100Kbeta}, there
is a  clear dependence of $\lvert\tau_{4.8}\rvert$ on the behaviour of the SED at
frequencies lower than the turnover frequency:  flatter power laws (smaller $p$) result in
smaller values of $\lvert\tau_{4.8}\rvert$ compared to steeper power laws (larger $p$).

As a last result we note that {by following exactly the same calculational procedure
as when collisions are included, no evidence could be found for the \gstate{} transition
being inverted by the external dust radiation field (i.e. when collisions are switched
off).} 
\section{Discussion}

Inspection of Figs.\,\ref{fig:taucollonly} to \ref{fig:tautdgb100Kbeta} shows some
interesting behaviour of the \gstate{} maser which had not been seen or anticipated in the
results of \citet{vanderwalt2014}. Although the basic conclusions of
\citet{vanderwalt2014} are, in a limited sense, still valid, this study shows that the
presence of a far-infrared dust radiation field significantly affects the region in the
\nhtwo{}-\tk{} plane where an inversion can occur. This obviously calls into question the
conclusion by \citet{vanderwalt2014} that the 4.8 GHz \formaldehyde{} masers are strictly
collisionally pumped. If a far-infrared radiation field has such a significant effect on
the 4.8 GHz masers, then what is the role of collisions, especially if no inversion seems
to occur when collisions are switched off?  There are two theoretical limiting results
which indicate that collisions play a central role in the inversion of the \gstate{}
maser. The first is that an inversion can be obtained without the presence of an external
far-infrared radiation field, and the second that no inversion is found if collisions are
switched off even if an external far-infrared radiation field is present. These two
limiting cases, as well as the other results, must be the consequence of a specific pumping
scheme for the formaldehyde masers. We propose a possible pumping scheme and attempt to
explain the results within this framework.

\subsection{Proposed pumping scheme}
To create a population inversion for the \gstate{} transition, it is necessary that the
excitation rate out of the $1_{1\,1}$ state be faster than the rate at which the
$1_{1\,0}$ state is populated through downward transitions from higher energy
states. Inspection of the Einstein A coefficients shows that the fastest radiative
downward route to $1_{1\,0}$ is through the upper levels of the doublet states of $K_a =
1$ (e.g. $7_{1\,6}\rightarrow 6_{1\,5} \rightarrow 5_{1\,4} \rightarrow 4_{1\,3}
\rightarrow 3_{1\,2} \rightarrow 2_{1\,1} \rightarrow 1_{1\,0}$). It therefore seems  that
the pumping scheme must be such that there is a route and mechanism whereby molecules can
be excited to the upper levels of the doublet states. Direct upward radiative
excitation out of $1_{1\,1}$ to the upper levels (even $K_c$ values) of the doublet states
is limited to $\Delta J = 0,1,~\Delta K_c = \pm 1$ and must therefore start with
excitation first to $1_{1\,0}$. However, this transition is very slow. For example, for an
undiluted dust radiation field as in Eq.\,\ref{eq:dustsed} with \td{} = 100 K and $p =
1.15$ (the case for Arp 220; the exact value is not important for this example), the
radiative transition rate for $1_{1\,1} \rightarrow 1_{1\,0}$ is only $7.4 \times
10^{-10}~\mathrm{s^{-1}}$. On the other hand, the corresponding rate for $1_{1\,1}
\rightarrow 2_{1\,2}$ is $2.9 \times 10^{-5}~\mathrm{s^{-1}}$.  Upward radiative
excitation out of $1_{1\,1}$ via $2_{1\,2}$ to the lower levels of higher $J$ doublet
states (e.g. $3_{1\,3},~4_{1\,4},~5_{1\,5}$) is therefore much faster than via
$1_{1\,0}$ to the upper levels of higher $J$ doublet states
(e.g. $2_{1\,1},~3_{1\,2},~4_{1\,3},~5_{1\,4}$).

To populate the $1_{1\,0}$ state, and since radiative excitation out of $1_{1\,1}$ via
$1_{1\,0}$ to the upper levels of the doublet states is very slow, it is necessary to
transfer molecules from the lower levels of the doublet states to the upper levels of the
doublet states, which then provides a fast radiative decay route to $1_{1\,0}$.  However,
radiative transitions from the lower levels of the doublet states to the associated upper
levels are very slow compared to radiative transitions between the lower levels of the
doublet states. For example, the radiative transition rate for $5_{1\,5} \rightarrow
5_{1\,4}$ is $6.2 \times 10^{-8}\,\mathrm{s^{-1}}$ for \td{} = 100 K and $p = 1.15$. On
the other hand, the radiative transition rate for $5_{1\,5} \rightarrow 4_{1\,4}$ is $1.2
\times 10^{-3}~\mathrm{s^{-1}}$ and $4.2 \times 10^{-4}~\mathrm{s^{-1}}$ for $5_{1\,5}
\rightarrow 6_{1\,6}$.  The implication is that if a molecule is excited from $1_{1\,1}$
to $2_{1\,2}$ (the lower doublet state for $J = 2$), further upward and downward radiative
excitations will   only be within the ladder of lower level doublet states.

\begin{figure}
  \centering
  \includegraphics[scale=0.65]{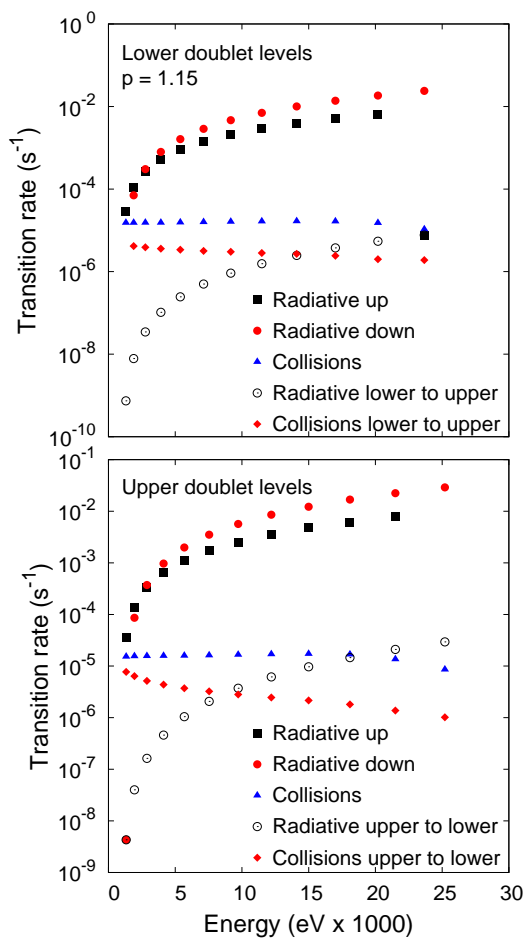}
  \caption
      { Radiative and collision rates  of the different energy levels for
        the lower levels of the doublet states (upper panel) and upper levels of the
        doublet states (bottom panel). The rates were calculated for \td{} = 100 K, \tk{} = 180
        K, \nhtwo{} = $10^{4.25}~\mathrm{cm^{-3}}$, and $p = 1.15$. Radiative up (black
        squares) means upward  radiative excitation out of a state (absorption); radiative
        down (red filled circles) means the sum of spontaneous and stimulated emission rates;
        collisions (blue triangles) means total collision rate out of a state; radiative
        lower to upper (open circles) means radiative rate for absorption from the lower
        level of a doublet state to the associated upper level; collisions lower to upper
        (red diamonds) means the collision rate from the lower level of a doublet state to
        its associated upper level.
      }
  \label{fig:p1p15rates}
\end{figure}

We illustrate the above in Fig.\,\ref{fig:p1p15rates}, where radiative and collisional
rates for different transitions are compared for the case when \nhtwo{} =
$10^{4.25}~\mathrm{cm^{-3}}$, \tk{} = 180 K, \td{} = 100 K, and $p = 1.15$. In the upper
panel we show the radiative rates for upward transitions ($\Delta J = + 1$, $\Delta K_c =
+1$, black squares) between the lower levels of the doublet states, the corresponding
downward rates ($\Delta J = -1$, $\Delta K_c = -1$, filled red dots), and the upward
radiative rates from the lower levels of the doublet states to the corresponding upper
levels ($\Delta J = 0, ~\Delta K_c = -1$, e.g. $5_{1\,5} \rightarrow 5_{1\,4}$, open
circles) for $K_a = 1$. Radiative rates between the $K_a = 1$ and $K_a = 3$ ladders are so
low that radiative transitions between the ladders can be ignored. Also shown are the
total collision rates out of the lower level of the doublet states to any other level
(blue triangles), and more specifically the collision rates from the lower levels of the
doublets to any of the upper level doublet states (red diamonds). As can be seen, the
upward radiative rates between the lower levels of the doublet states (black squares) are
orders of magnitude higher than the radiative rates from the lower levels of the doublet
states to their associated upper levels of the doublet states (open circles). This
illustrates the point made above that if a molecule is excited (radiative or
collisionally) from $1_{1\,1}$ to the $2_{1\,2}$ state, further radiative transitions will
only be within the ladder of the lower levels of the doublet states and that radiative
transfer of molecules from the lower levels of the doublet states to their associated
upper levels is not effective.  On the other hand, under the given physical conditions,
collision rates from the lower levels of the doublet states to any of the upper levels of
doublet states (red diamonds) are significantly higher than the corresponding radiative
rates.  This is also the case, at least for the seven lowest levels of the doublet states
above $1_{1\,1}$, specifically for collisional transfer of molecules from the lower levels
of the doublet states to any of the upper levels of the doublet states (red diamonds). It
can therefore be concluded that collisions are the mechanism responsible for population
transfer from the lower to the upper levels of the doublet states.

The bottom panel of Fig.\,\ref{fig:p1p15rates} shows the corresponding case for the upper
levels of the doublet states. Also in this case it is seen that the radiative rates for
transitions from the upper levels of the doublet states to the associated lower levels is
small compared to radiative rates between the different upper levels of the doublet states
and the collisional rates  from the upper to the lower levels of the doublet
states. Both the upper and lower panels show rather clearly that population exchange between
the lower and upper levels of the doublet states is preferentially through collisions and
not radiation.  

\begin{figure}
  \centering
  \includegraphics[scale=0.65]{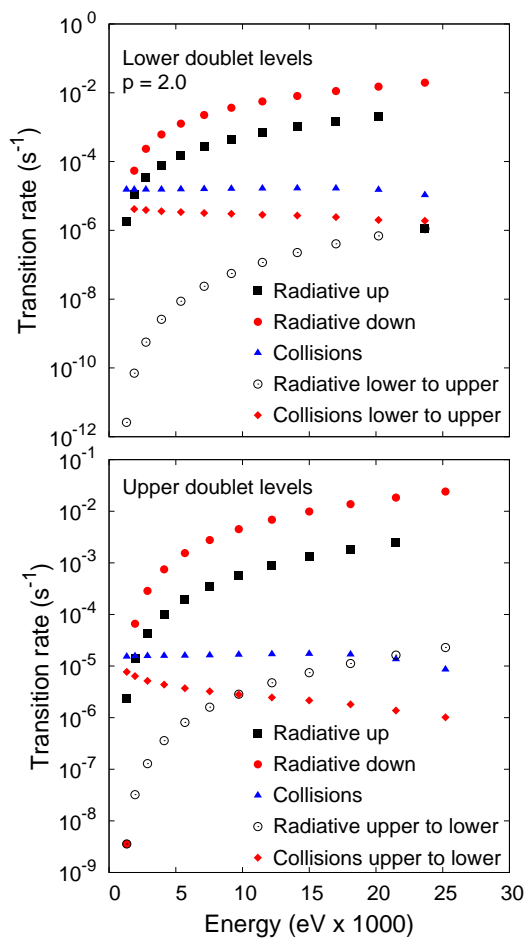}
  \caption
      {
        Same as for Fig.\,\ref{fig:p1p15rates}, but for $p = 2.0$.
      }
  \label{fig:p2p0rates}
\end{figure}

Excitation out of $1_{1\,1}$ need not be mainly radiative, but for a given \tk{} and
\nhtwo{} it also depends on the energy density in the radiation field at the respective
frequencies for the $1_{1\,1} \rightarrow 1_{1\,0}$ (4.8 GHz) and $1_{1\,1} \rightarrow
2_{1\,2}$ ($\sim$140 GHz) transitions. In Fig.\,\ref{fig:p2p0rates} we show the case when
$p = 2$, in other words when the SED for $\nu \ll \nu_0$ (Eq.\,\ref{eq:dustsed}) is
significantly steeper ($I_\nu \propto \nu^{4}$) than for $p = 1.15$ ($I_\nu \propto
\nu^{3.15}$).  It is seen that the upward radiative rates for the lower levels of the
doublet states (upper panel, black squares) are significantly lower and for the lower $J$
levels even lower than the collisional rates. Collisional excitation out of $1_{1\,1}$
actually dominates radiative excitation in this case. The radiative rates for the transfer
of molecules from the lower levels of the doublet states to any of the upper levels of the
doublet states (open circles) are now also significantly smaller compared to the
corresponding collision rates (red diamonds). Thus, as for $p = 1.15$, population exchange
between the lower and upper levels of the doublet states through collisions is by
far the dominant process for $p = 2$.

\begin{figure}
  \centering
  \includegraphics[scale=0.65]{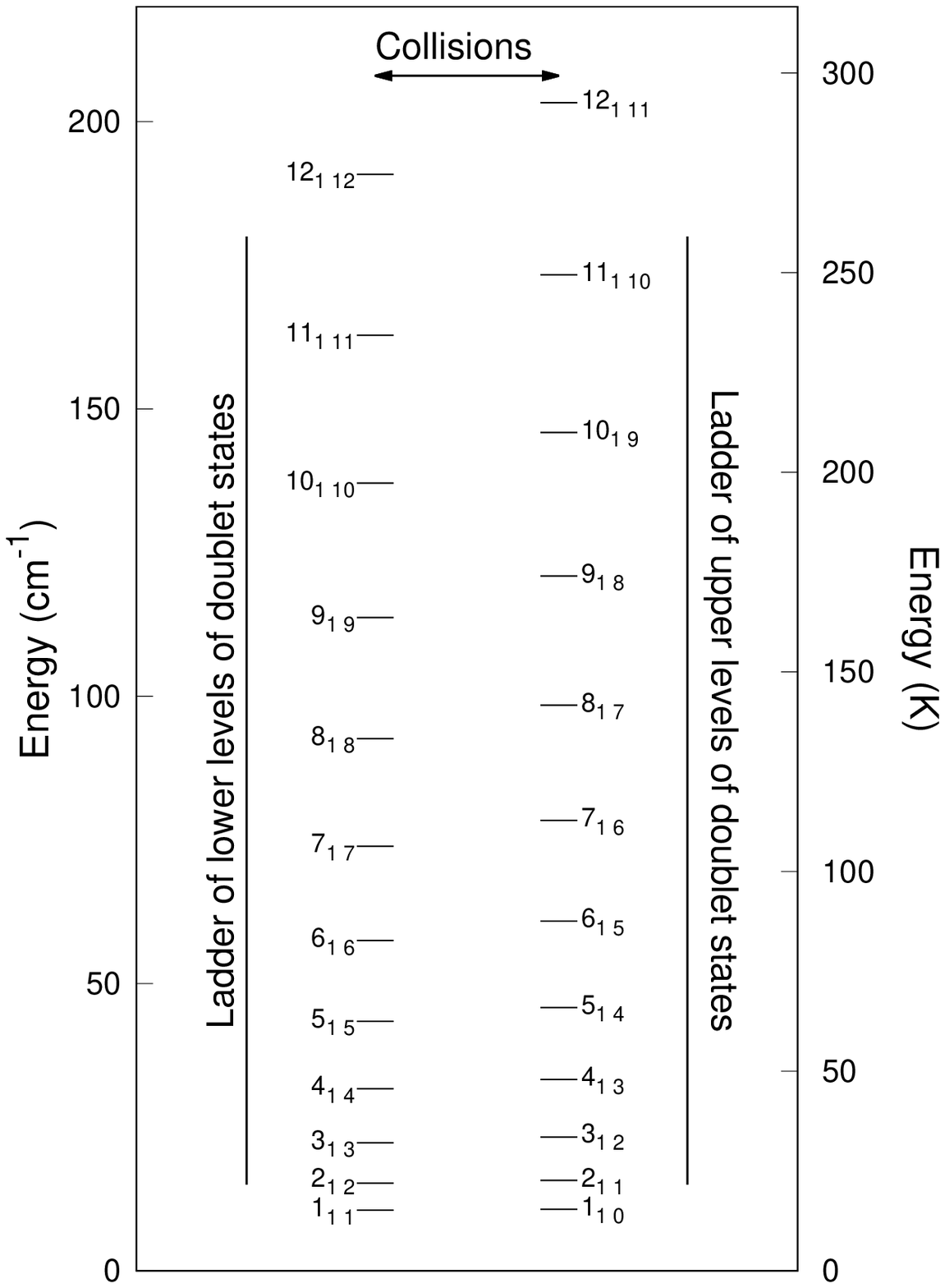}
  \caption
      { Energy level diagram of lower levels of the doublets and the upper levels of the
        doublets to illustrate the proposed pumping scheme for the 4.8 GHz \formaldehyde{}
        masers. Collisions couple the two sets of states.}
  \label{fig:pumping}
\end{figure}

A possible pumping scheme is therefore that molecules are radiatively or collisionally
excited out of the $1_{1\,1}$ state. Radiative excitation out of $1_{1\,1}$ is almost
exclusively to the $2_{1\,2}$ state from where further upward radiative excitations to the
lower levels of the doublets can take place. Collisional excitation out of $1_{1\,1}$ can
be either to the lower or upper levels of the doublets. Radiative excitation out of
$1_{1\,1}$ to $1_{1\,0}$ is very slow, which means that the ladder of lower level doublet
states will be populated significantly faster than the ladder of upper level doublet
states. A fraction of the population in the lower level doublet states is collisionally
transferred to the upper level doublet states, which then provides a relatively fast
downward radiative route to $1_{1\,0}$ to create a population
inversion. Figure\,\ref{fig:pumping} is a schematic representation of the proposed pumping
scheme.

The above-proposed pumping scheme can account, at least qualitatively, for the presented
results. First, it is clear that without collisions (i.e.  with only the internal and
external radiation fields present) there can be no effective population transfer from
the ladder of lower doublet states to the upper doublet states to eventually overpopulate
the $1_{1\,0}$ state relative to the $1_{1\,1}$ state. This explains the absence of an
inversion when collisions are switched off. Comparison of the bottom panel of
Fig.\,\ref{fig:tautdgb100K} and both panels of Fig.\,\ref{fig:tautdgb100Kbeta} shows that
the inversion of the \gstate{} transition is smaller for $p = 0.5$ compared to the cases
of $p = 1.15$ and $p = 2$. Since the energy density in the radiation field varies as
$\nu^{2+p}$ for $\nu \ll \nu_0$, it follows that for $p = 0.5$ the radiative excitation
rates out of the lower levels of the doublets is significantly higher than for $p =
2$. The upper panel of Fig.\,\ref{fig:p1p15rates} also shows that the upward and downward
radiative rates for the lower levels of the doublet states are very similar for
$p=1.15$. As a consequence, molecules are more or less `trapped' within the ladder of
lower doublet states when $p = 0.5$ with the probability to be collisionally transferred to
the ladder of upper levels of the doublet states being small. On the other hand, for $p =
2$ the downward rates within the ladder of lower level doublet states are markedly higher
than the upward rates, meaning that molecules in higher $J$ states will rather quickly
radiatively decay to lower states where the probability for collisional transfer to the
ladder of upper level doublet states is comparable to that of upward radiative excitation
within the ladder of the lower levels of the doublet states. Such a scenario can explain
why the inversion is larger for $p = 2$ than for $p = 0.5$.

\subsection{Constraints due to the \formaldehyde{} abundance}
\label{sec:abundances}
\begin{figure}
  \centering
  \includegraphics[scale=0.65]{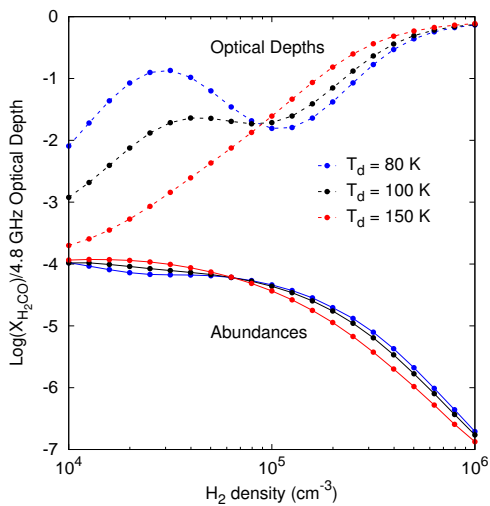}
  \caption
      { Plot showing $X_{\mathrm{H_2CO}}$(solid lines) and  4.8 GHz optical depth (dashed
        lines) as a function of \htwo{} density for \tk{} = 180 K, $p = 1.15$, and $W = 1$
        for different dust temperatures.  A maser path length of $10^{17}$ cm and line width
        of 1 $\mathrm{km\,s^{-1}}$ was used to calculate the \formaldehyde{}
        abundance.}
  \label{fig:abundances}
\end{figure}

Figures \ref{fig:tautdgb80K} to \ref{fig:tautdgb100Kbeta} suggest that inversion of the
\gstate{} transition can happen over a significant part of the parameter space with the
implication that the 4.8 GHz masers should be quite abundant. However,  the
\formaldehyde{} masers are very rare, which means that there must be other factors that affect
the occurrence of these masers. \citet{vanderwalt2014} noted that his results imply that
the \formaldehyde{} abundance must be significantly higher than what is generally
observed. In our calculations the \formaldehyde{} abundance, $X_{\mathrm{H_2CO}}$, is
implicitly included in the specific column density given by $X_{\mathrm{H_2CO}}
n_{\mathrm{H_2}} \ell/\Delta \varv$, where $\ell$ is the maser path length and $\Delta
\varv$ the maser line width. In the variation of $\tau_{4.8}$ with the specific column
density (e.g. as in Fig.\,\ref{fig:baancase}) the most obvious specific column density
that can be used to calculate the abundance is that where the optical depth has its
maximum negative value. Figure \,\ref{fig:abundances} shows, as an example, the variation
of the optical depth and $X_{\mathrm{H_2CO}}$ at \tk{} = 180 K for \td{} = 80, 100, 150 K,
W = 1 (Figs.\,\ref{fig:tautdgb80K} - \ref{fig:tautdgb100Kbeta}), $\Delta \varv =
10^5~\mathrm{cm\,s^{-1}}$, and $\ell = 10^{17}$ cm \citep{Cragg2002}. The variation of the
optical depth with \nhtwo{} for the three dust temperatures is seen to be quite
different. On the other hand, the derived \formaldehyde{} abundances behave in the
same way and vary between $\sim 3.94 \times 10^{-3}$ and $\sim 6.87 \times 10^{-6}$ from
\nhtwo{} = $10^4~\mathrm{cm^{-3}}$ to $10^6~\mathrm{cm^{-3}}$. Using the filling factor
corrected data presented in Table 8 of \citet{Ginsburg2011}, an average \nhtwo{} of $4.5
\times 10^4~\mathrm{cm^{-3}}$ and average \formaldehyde{} abundance of $1.1 \times
10^{-11}$ for more than 60 lines of sight were found. It is clear that the observed
abundances are orders of magnitude too small to explain the masers. The required
abundances are significantly higher than  most models can produce \citep[see
  e.g.][]{Guzman2011}. It is worth noting, however,  that \cite{Maret2004} derived a
\formaldehyde{} abundance of $6\times 10^{-6}$ for the warm inner envelope of the young
low mass star L1527. This suggests that similar cases may also exist for young high mass
stars.

\begin{figure}
  \centering
  \includegraphics[scale=0.57]{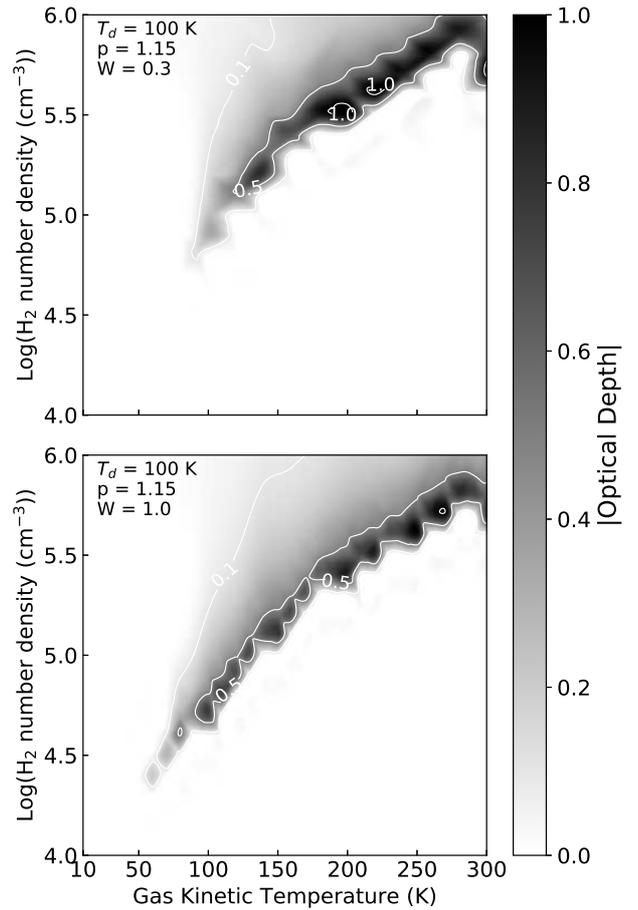}
  \caption
      { Variation of $\lvert\tau_{4.8}\rvert$ in the \nhtwo{}-\tk{} plane, with the SED of
        the dust emission   given by Eq.\,\ref{eq:dustsed}, with $p = 1.15$ and
        $\mathrm{T_d}$ = 100 K and when the \formaldehyde{} abundance is restricted to be
        less than $10^{-5}$. The upper panel is for W = 0.3 and the lower panel for W =
        1. This result should  be compared with that shown in Fig.\,\ref{fig:tautdgb100K}
        where no upper limit is placed on the \formaldehyde{} abundance.}
  \label{fig:xlimit}
\end{figure}

Recently \citet{Vichietti2016} proposed a chemical route for the production of
\formaldehyde{} in young high mass star forming regions leading to abundances as high as
$10^{-5}$.  Using this as a guide, we illustrate in Fig.\,\ref{fig:xlimit} what the effect
is of setting an upper limit of $10^{-5}$ for the \formaldehyde{} abundance on the region
in the \nhtwo{}-\tk{} plane where an inversion can occur. Comparison with
Fig.\,\ref{fig:tautdgb100K} shows that the allowed region for an inversion is
significantly smaller than when no constraint is put on the \formaldehyde{} abundance. Such
a significantly smaller region in the \nhtwo{}-\tk{} plane might be a reason for the
rarity of the \formaldehyde{} masers.

An important aspect of Fig.\,\ref{fig:xlimit} is that for W = 0.3 (upper panel) inversion
of the \gstate{} transition occurs basically for $n_{\mathrm{H_2}} >
10^{5.5}\,\mathrm{cm^{-3}}$. Within the framework of our calculations the implication is
that the collision rates as shown in Figs.\,\ref{fig:p1p15rates} and \ref{fig:p2p0rates}
(blue triangles and red diamonds) have to be increased by a factor of 1.25 dex. If this is
done the collisional excitation rate out of \zerostate{} is seen to dominate the radiative
excitation rate by more than an order of magnitude. If this is the case it can indeed be
said that the \formaldehyde{} masers are collisionally pumped. In fact, this explains why
there is a region in the \nhtwo{}-\tk{} plane where an inversion can occur without the
presence of an external radiation field.

\subsection{Evaluation of the results of \citet{Baan2017}}

As is shown in Section\, \ref{sec:baan}, using the same set of parameter values as was
used by \citet{Baan2017}, the population inversion became weaker when \tk{} approached
\td{} and disappeared completely when \tk{} = \td{}. As we  show in
Fig.\,\ref{fig:baanlevpops}, the disappearance of the population inversion is most
definitely due to the system being in thermodynamic equilibrium when \tk{} = \td{} and
that a population inversion is not possible in such a case. It can also be seen, especially
from Figs.\,\ref{fig:tautdgb100K} and \ref{fig:tautdgb150K}, that the condition $T_k <
T_d$ as a requirement for an inversion, as set by \citet{Baan2017} (and not by
\citealt{vanderwalt2014}, as stated by these authors), does not follow from our results. 

The results presented above have some definite implications for interpreting the
\formaldehyde{} megamaser emission presented by \citet{Baan2017}. It is first necessary
to note that no inversion is found for $T_k \lesssim$ 100 K when using a more realistic
SED with \td{} = 59 K for the dust emission (Fig.\,\ref{fig:tauarpdust}). A direct
implication of this result is that, contrary to the conclusions by \citet{Baan2017}, the
4.8 GHz \formaldehyde{} megamasers in the three starburst galaxies are not associated
with cold material. Furthermore, if the effect of an upper limit of $10^{-5}$ on the
abundance of o-\formaldehyde{} is taken into account (Fig.\,\ref{fig:xlimit}), the
implication is that the masers trace hot (100 - 300 K) gas with densities $10^5 -
10^6~\mathrm{cm^{-3}}$ rather than cold gas. Such kinetic temperatures have recently been
reported by \citet{Gieser2021} for quite a number of high mass star forming regions. It is
fairly well established that the 4.8 GHz masers in the Galaxy are associated exclusively
with high mass star forming regions \citep{Araya2015}.  There is thus no reason why it
should not also be the case for starburst galaxies. In the three starburst
galaxies observed by \citet{Baan2017} the 4.8 GHz maser emission is spatially extended and
associated with star forming regions, which by itself suggests the association of these masers
with high mass star forming regions as in the case of the Galaxy. Why \formaldehyde{}
maser emission is associated with only three starburst galaxies, and  exactly why it
is rare in the Galaxy, are   questions still to be answered.

\section{Conclusions}

 We performed a new set of numerical calculations
 to investigate the inversion of the \gstate{} transition of o-\formaldehyde{}. The
 calculations covered a significantly larger part of parameter space compared to previous
 calculations.  Considering the results presented above we summarize and conclude as
 follows:
\begin{enumerate}
\item While theoretically there is a population inversion of the \gstate{} transition when
  using a 50 K black-body radiation field, the decrease in the optical depth as the
  kinetic temperature is increased from 10 to 40 K, as presented if Fig. 19 of
  \citet{Baan2017}, is due to the system approaching thermodynamic equilibrium. This
  behaviour does not imply that the inversion is the consequence of far-infrared pumping.
  When a more realistic grey-body SED corresponding to that of
  Arp 220 is used, no inversion of the \gstate{} transition is found for \tk{} < 100
  K. Our calculations suggest that the megamasers in the three starburst galaxies observed
  by \citet{Baan2017} are not associated with cold material, but rather with dense and hot
  gas associated with high mass star formation.
\item We  show that, within the framework of our calculations, the \gstate{}
  transition can be inverted without the presence of a far-infrared radiation field. On
  the other hand, no evidence could be found for the \gstate{} transition being inverted
  by the external dust radiation field (i.e. when collisions are switched off). Collisions
  therefore seems to play a key role in the pumping mechanism of the masers.
\item We proposed a pumping scheme in which molecules are excited either radiatively or
  collisionally out of the $1_{1\,1}$ state. Radiative excitation out of $1_{1\,1}$ leads
  to the population of the ladder of lower levels of the doublet states. Population
  exchange between the ladders of lower and upper doublet states is collisional since
  radiative coupling between the two ladders is weak. At high \htwo{} densities
  collisional excitation out of $1_{1\,1}$ can be to both ladders. Molecules excited
  collisionally to higher states in the ladder of lower levels of the doublet states still
  have to be transferred collisionally to the ladder of upper levels of the doublet
  states. Without collisions, population exchange between the ladders of the lower and
  upper levels of the doublet states is extremely slow. Transfer of molecules from the
  lower levels of the doublet states to the upper levels of the doublet states is
  necessary since it then provides a fast radiative decay route to $1_{1\,0}$.
\item When assuming reasonable values for the maser line width and the maser path length it
  is found that, over most of parameter space where theoretically an inversion can occur,
  the required abundance of \formaldehyde{} is significantly larger than observed and
  predicted by models. Limiting the \formaldehyde{} abundance to less than $10^{-5}$
  significantly reduces the region in the \nhtwo{}-\tk{} plane where an inversion can
  occur to a rather narrow region with \htwo{} densities greater than $\sim
  10^5~\mathrm{cm^{-3}}$ and \tk{} $\gtrsim$ 100 K. This might explain the rarity of the
  \formaldehyde{} masers in the Galaxy and starburst galaxies. 
\end{enumerate}

We finally point out that o-\formaldehyde{} has several submillimeter transitions between
excited states (for which $E_{upper} > 100$ K) which can be used to probe the physical
conditions of the environments of the \formaldehyde{} masers.  Dedicated high resolution
observations of these transitions can be used to test the predictions of our model
calculations.

\begin{acknowledgements}
      This work is based on research supported in part by the National Research Foundation
      of South Africa (Grant Numbers 132494).
\end{acknowledgements}

\bibliographystyle{aa}
\bibliography{ref}

\end{document}